# Ultra-broadband, few cycle pulses directly from a Mamyshev fiber oscillator

Chunyang Ma[1,2], Ankita Khanolkar[3], Yimin Zang[3], and Andy Chong [2,3]*

1.Department of Electronic Science and Engineering, Jilin University, Changchun, Jilin Province, 130012, China
2.Department of Physics, University of Dayton, Dayton, OH 45469, USA
3.Department of Electro-Optics and Photonics, University of Dayton, Dayton, OH 45469, USA
*Corresponding author: achong1@udayton.edu

While the performance of mode-locked fiber lasers has been improved significantly, the limited gain bandwidth restricts them to generate ultrashort pulses approaching a few cycles or even shorter.  Here we present a novel method to achieve few cycle pulses (~5 cycles) with ultra-broad spectrum (~400 nm).  To our best knowledge, this is the shortest pulse width and broadest spectrum directly generated from fiber lasers.  It is noteworthy that a dramatic ultrashort pulse evolution can be stabilized in a laser oscillator by the unique nonlinear processes of a self-similar evolution as a nonlinear attractor in the gain fiber and a perfect saturable absorber action of the Mamyshev oscillator.

# Introduction

Ultrafast optics has been led by bursts of optical pulses from mode-locked lasers. Such ultrafast pulses have been as short as the femtosecond ($10^{-15}$ second) scale and even down to attosecond ($10^{-18}$ second) scale recently. They have enabled key scientific researches such as femtochemistry [1] and the optical frequency comb [2, 3] which received Nobel prizes in 1999 and 2005 respectively. Not only frontier researches enabled by ultrafast pulses but also generation of short, powerful optical pulses itself has been recognized as intriguing science to receive a Nobel prize in 2018 [4].

Ultrafast mode-locked lasers have enabled many broader impact applications such as nonlinear microscopy [5], materials processing [6], surgical applications [7], sensing [8] etc. For those applications, the demand for shorter and higher peak power pulses has steadily grown up. Particularly, optical pulses of temporal durations reaching few cycles and even a single cycle of the carrier frequency with broad spectra possibly even beyond an octave spanning have been essential for many applications. Such short pulses are suitable for attosecond science [9,10], high harmonic generation [11], coherent X-ray generation [12], etc. A popular and straightforward method to obtain such short pulses is the nonlinear pulse compression technique. The spectrum of a high peak power pulse can be broadened in a nonlinear media with some nonlinear phase accumulation. By compensating the nonlinear phase properly, pulses can be compressed to much shorter durations than those directly from mode-locked lasers [13-16].

Even though the pulse compression technique is well established, generating short pulses with broad spectra generated directly from mode-locked lasers have advantages. In a laser cavity, the temporal phase on a pulse is to be reshaped for each round trip. Therefore, pulses directly from a laser are to be more stable with less noise. Furthermore, pulses directly from a laser cavity can be conveniently compressed externally for even shorter pulses with broader spectra. Solid state lasers, such as a Ti:sapphire laser, have been noticeably successful in generating short pulses reaching few cycles directly from a laser oscillator [17-19]. By a careful intracavity dispersion compensation of the gain crystal, 5 fs pulses, which is shorter than two cycles with octave spanning spectra, have been directly generated from a mode-locked Ti: sapphire laser [19].

Recently, the demand for mode-locked fiber lasers is steadily rising with more and more fiber laser products emerging in the market. Even though solid state lasers have been successful in many applications, they are bulky, high-cost, and not use friendly in general. In contrast, mode-locked fiber lasers have some practical advantages of compactness, low-cost, efficient heat control, and environmental stability. In parallel, the performance of fiber lasers has been improved steadily in pulse energies and peak powers to be suitable for broader industrial applications [20].

However, pulse durations and spectral bandwidths (BW) from fiber lasers still fall well behind those of solid state counterparts. It has been extremely challenging to generate short pulses approaching a few and a single cycle in fiber devices. The main obstacle for short pulses from fiber lasers is the limited gain BW. For example, the Ti:sapphire crystal has a very wide emission spectral range from 650 nm to 1100 nm which is proper to generate few cycle pulses. Meanwhile, the broadest gain fiber BW is ~40 nm of Ytterbium (Yb) doped fibers which corresponds to ~ 30 fs pulse duration.

To reach shorter than 30 fs in fiber devices with limited gain BW, the only reliable method for substantial spectral broadening beyond the gain BW is a strong nonlinear spectral broadening by the self-phase modulation (SPM). For example, a single cycle pulse has been generated from a fiber device by coherently combining two extremely broadened spectra in highly nonlinear photonic crystal fibers (PCF) [21]. Even though the method was successful, such devices are way too complex with possible stability issues. Therefore, it is strongly desirable to generate broad spectra beyond the gain BW from a simple fiber oscillator. To generate such a broad spectrum directly from a fiber oscillator, a meticulous and possibly a brand new laser design is crucial to permit a substantial intracavity spectral broadening without losing the mode-locked operation.

It turns out that a recently discovered laser design facilitates such laser operation. In 2010, a new pulse shaping mechanism referred as self-similar pulse evolution was demonstrated in mode-locked fiber lasers [22-24]. Self-similar pulses are linearly chirped parabolic pulses as asymptotic solutions of a normally dispersive fiber amplifier [25]. Remarkably, a self-similar pulse is a strong nonlinear attractor of the gain fiber. Even though a strong perturbation such as an accumulation of nonlinear phase is introduced, the perturbed pulse quickly comes back to an attractor which is a self-similar parabolic pulse [26]. As a consequence, an enormous intracavity spectral broadening can be stabilized as a mode-locked operation by a strong spectral filtering and the self-similar amplification. Such an idea was first implemented in 2012 for Yb-doped self-similar fiber lasers. By inserting a highly nonlinear PCF after the self-similar pulse is established in the gain fiber, a very broad spectrum (~200 nm BW), which is much larger than the gain BW, has been demonstrated. Broad spectrum can be spectral filtered and comes back to a self-similar pulse again as an attractor of the gain fiber. Pulses from the laser are highly chirped but could be dechirped externally to ~20 fs [27]. It is noteworthy that a new pulse shaping mechanism by self-similar evolution permits a mode-locked spectrum much larger than the gain BW.

Since then, there hasn't been much improvement of the pulse duration from fiber lasers. The main problem is that the peak power of the chirped parabolic pulse into the PCF could not be improved significantly. Of course, to generate even broader spectra with shorter pulse durations, high peak power in the PCF is desirable. It has been recently suggested that an intracavity dispersion delay line (DDL) can increase the pulse peak power by reducing the chirp on the pulse. Therefore, a significant spectral broadening can occur in the PCF. Although the numerical simulation indicates that even an octave spanning spectrum is possible in this design [28], such broad spectra have not been observed in experiments. One plausible conjecture is the non-ideal saturable absorber in real lasers. For example, a very strong close to an ideal saturable absorber was assumed while saturable absorbers in real lasers are not ideal.

In 2017, self-similar amplifiers with a step-like saturable absorber called Mamyshev oscillator has been demonstrated [29]. Mamyshev oscillator shows a transmission – intensity curve that jumps from zero to one like a step-function at a certain intensity, which is referred as a 'perfect' saturable absorber [30]. A Mamyshev oscillator already demonstrated its exceptional potential for high pulse energies (50 nJ) and short-pulse (40 fs) durations [29]. Recently, it achieved very noticeable performance of pulse energy 1 µJ and 40 fs pulse duration [31]. Owing to its unique 'perfect' saturable absorbing action, the Mamyshev oscillator appears to be a good candidate to stabilize intracavity spectrum broadening.

Here, we present the generation of few cycle pulses directly from a mode-locked Mamyshev fiber oscillator. To enhance the intracavity spectral BW, a highly nonlinear PCF was inserted in the cavity. A very broad mode-locked spectrum (~400 nm) is generated. It is noticeable that such a broad spectrum broadening can be mode-locked owing to nonlinear processes such as a strong saturable absorber and the self-similar evolution. The output pulse was dechirped down to ~5 cycles. To our best knowledge, this is the broadest spectrum (~400 nm) with the shortest pulse duration (17 fs) directly generated from a mode-lock fiber oscillator.

## Results

### Laser setup

The experimental set up of the Mamyshev oscillator is schematically illustrated in Fig. 1. The oscillator is consists of two fiber arms. In each arm, a pump combiner couples the 976 nm pump into the 3.1 m Yb-doped double cladding (DC) gain fiber (Nufern: SM-YDF-5/130-VIII), which is followed by a 0.7 m single mode fiber (SMF). The total fiber length is 9 m which corresponds to the total cavity dispersion of 0.146 ps$^2$. Two free space isolators ensure a unidirectional laser operation. Two 600 lines/mm gratings with collimators serve as narrow 4 nm BW Gaussian spectral filters centered at ~1040 nm and ~1050 nm wavelength while the center wavelength of the laser is at ~1045 nm. Two half wave plates (HWPs) after the isolators adjust the polarization to maximize diffraction efficiency of the diffraction grating.

The first arm is used only to generate self-similar operation in the gain fiber. The spectral BW is same as the pure self-similar evolution in the gain fiber. However, in the second arm, more components are inserted to enhance the intracavity spectral BW. After the second arm, an anomalously DDL of a 1000 lines/mm grating pair is inserted to compress the pulse. The compressed pulse is coupled into 45 cm of an all normal dispersion (ANDi) PCF with a 50% coupling efficiency. After the PCF, two quarter wave plates (QWPs), a HWP and a polarization beam splitter2 (PBS) serve as a nonlinear polarization evolution based artificial saturable absorber to start mode-locking.

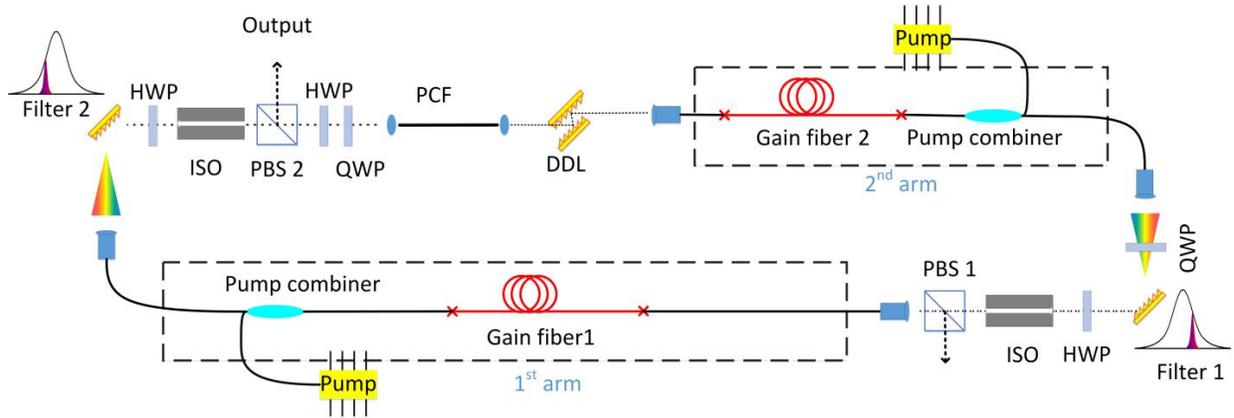

**Figure 1**. Schematic of the laser cavity. PBS: polarizing beam splitter. HWP: half wave plate. QWP: quarter wave plate. ISO: free space isolator. DDL: dispersion delay line (a grating pair), PCF: photonic crystal fiber.

## Numerical simulations

To understand the pulse evolution in the oscillator, we performed numerical simulations by the standard split-step Fourier method including Kerr nonlinearity, Raman effect, and up to fourth order group velocity dispersion. The simulation results are summarized in Fig. 2.

Fig.2 (a) shows the evolution of the pulse duration and the root mean squared (RMS) spectral BW over a round trip in the cavity. In the first arm, pulse duration and spectral BW do not change much initially in the passive SMF. However, the duration and the BW start to grow monotonically in the gain fiber and second passive SMF to form a positively chirped pulse. The spectral broadening in the gain fiber is clearly due to the self-similar evolution. Fig.2 (b) shows the evolution of the misfit parameter M. The misfit parameter, which indicates how much the pulse profile deviates from a parabola, is defined by $M^2 = \int (I - I_{fit})^2 / \int I^2 dt$, where $I$ is the pulse intensity profile while $I_{fit}$ is the parabolic pulse intensity profile. For example, $M = 0.14$ is for a Gaussian pulse, while $M \leq 0.06$ corresponds to a parabolic pulse [33]. Fig 2 (b) shows that the initial Gaussian pulse evolves to a parabolic pulse while propagating in the gain fiber. A parabolic pulse with increasing spectral BW is a key signature of the self-similar pulse evolution in normal dispersion gain fibers [22, 23]. The broadened self-similar spectrum is filtered by a sharp spectral filter (4 nm BW at 1050 nm) before the second arm.

In the second arm, the spectrum and the pulse durations broaden again in the gain fiber by the self-similar evolution but the DDL reduces the pulse duration significantly from 3.4 ps to 260 fs (Fig. 2(c)). Therefore, the pulse peak power is noticeably enhanced by ~13 times. The compressed pulse is coupled into the PCF which an ANDi PCF with a small positive GVD value and a very small mode field diameter (MFD). In the PCF, the combination of the high peak power, small MFD and small normal GVD broadens the spectral RMS BW significantly in the PCF. Over one round trip, the RMS spectral BW broadens from 3.44 nm to 175.4 nm corresponding a ~51 spectral breathing ratio. The broadened spectrum is coupled out of the cavity as an output. Fig. 2(d) and (e) show the output spectrum and pulse which is positively chirped. The spectrum shows 475 nm BW at -20dB. Fig. 2(f) shows the pulse profile (10 fs) after dechirped by a 300 lines/mm grating pair. Even though the full-width half-maximum (FWHM) pulse duration is close to the Fourier transform limited pulse duration (~9 fs), a long pedestal is observed due to uncompensated higher order dispersion effect. After the output is coupled out of the cavity, a sharp spectral filter (4 nm BW at 1040 nm) again cuts down the spectrum before the pulse returns to the first arm.

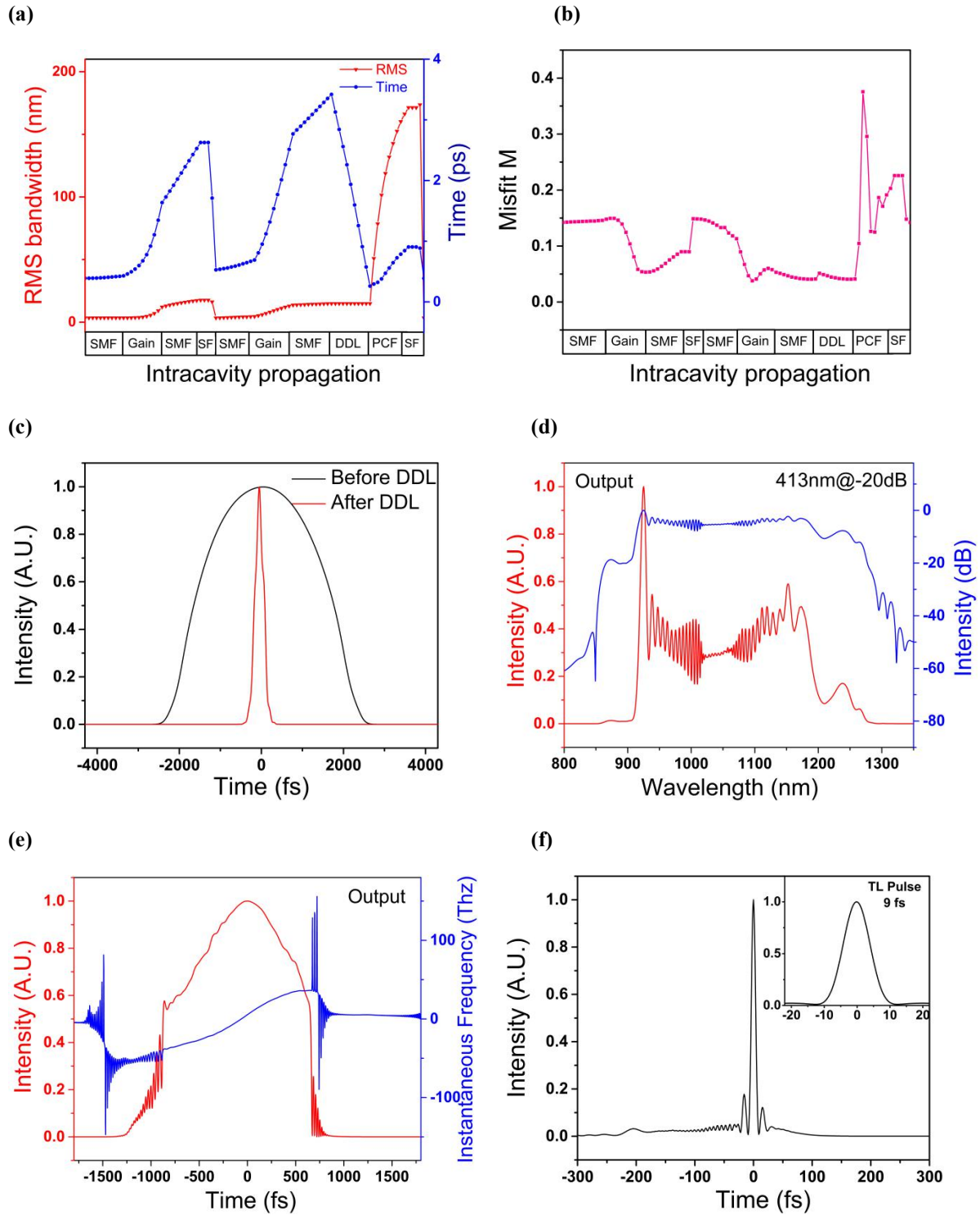

**Figure 2**. (a) Pulse duration and spectral BW evolution in the laser cavity. (b) Misfit parameter M in the cavity. (c) Pulse before and after the DDL. (d) Output spectrum. (e) Output pulse. (f) Numerically dechirped pulse by a 300 lines/mm grating pair.

**Experimental result**

The mode-locking is initiated by adjusting wave plates to induce the NPE based saturable absorber action.   Mode-locked results are presented in Fig. 3.   Fig. 3(a) shows the ejected spectrum at PBS1 while Fig. 3(b) shows the output spectrum at the PBS2, which is the enhanced spectrum after the PCF, with 394 nm BW at -20dB which is the broadest spectrum directly observed ever form mode-locked fiber lasers.   The output pulse energy is 3.5 nJ at 17.5 MHz repetition rate. The autocorrelation (Fig. 3(c)) shows 17 fs (~5 cycles) pulse duration after dechirped by a 300 /mm grating pair.   In the autocorrelation signal, noticeable pedestals are observed owing to uncompensated higher order group velocity dispersions by a grating pair.   It is believed that Fourier transform limited pulses (~9 fs, Fig. 3(d)) can be obtained with a better phase compensation technique such as a multiphoton intrapulse interference phase scans (MIIPS) [34].

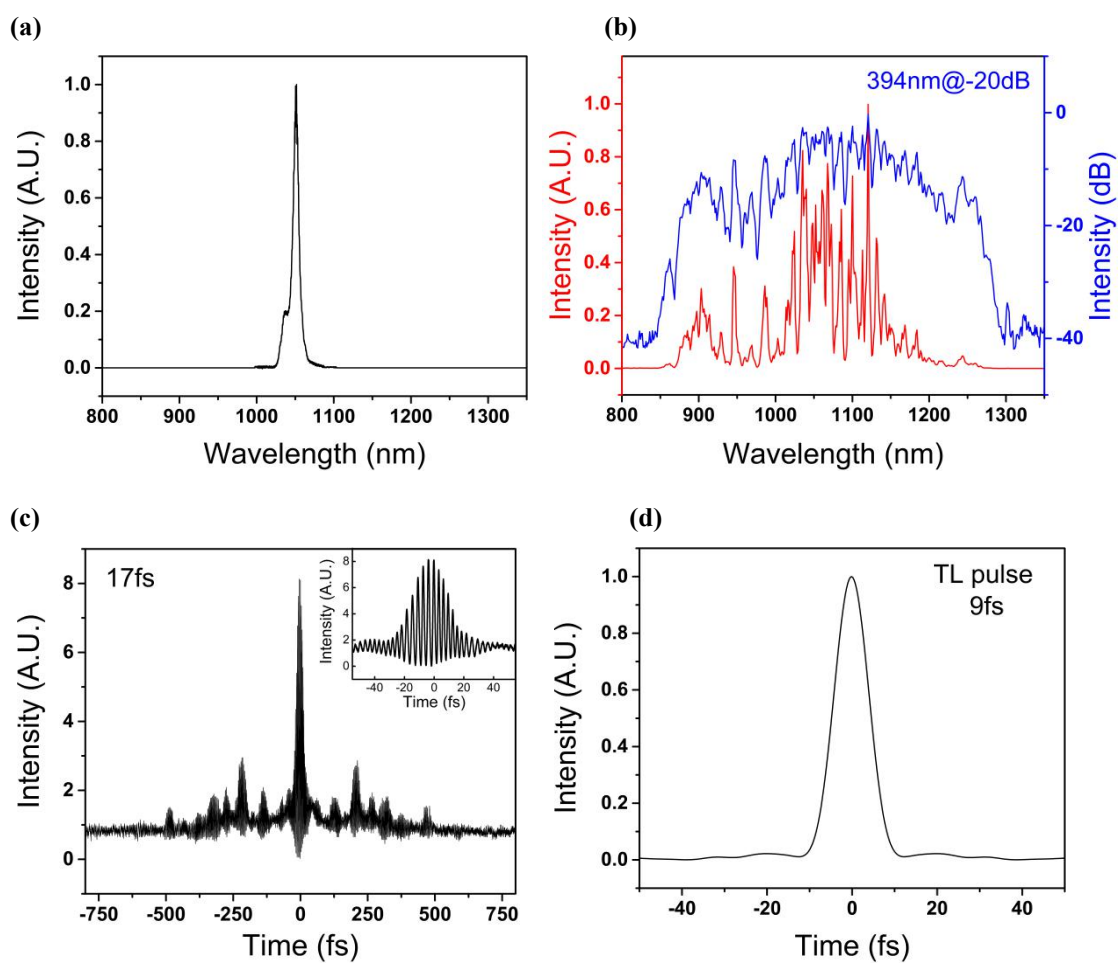

**Figure 3**. Experimental results. (a) Output spectrum from the PBS1. (b) Output spectrum. (c) Dechirped autocorrelation. (d) Calculated Fourier transform limited pulse profile.

**Discussion.**

In experiment, an anomalous DDL and a highly nonlinear ANDi PCF induced a huge spectral broadening. Of course, such an explosive spectral broadening in the cavity disturbs stable mode-locked operations. As mentioned in the introduction, there are two main mechanisms to stabilize the laser operation. First, there is a strong nonlinear attractor action which is a combination of the self-similar evolution in the gain fiber and a narrow spectral filtering. Ultrabroad spectrum is cut by a sharp spectral filter significantly but the BW grows again as it reaches the nonlinear attractor of the parabolic self-similar pulse. The nonlinear attraction action not only re-establishes a parabolic pulse profile but also cleans the complex nonlinear phase shift. Second is the 'perfect' step like saturable absorption action of the Mamyshev regenerator. In fact, numerical simulations indicate that it is possible to obtain such a broad spectrum in a simple self-similar laser [28]. However, the simulation could not generate such a broad spectrum unless an ideal saturable absorber with a very strong modulation depth (nearly 100%) is assumed in simulations. Therefore, it will be very challenging to generate such saturable absorber conditions experimentally. However, it is verified in numerical studies and experiments that the Mamyshev oscillator's step-like saturable absorber is crucial to stabilize such broad spectra. Numerical studies also show that we can generate even an octave spanning spectrum ( 744 nm at -20 dB) directly from the Mamyshev oscillator, the TL pulse width is just 5 fs which is nearly to a single cycle pulse (Fig. 5 (a) and (b)). It means it is possible to generate nearly single cycle pulses directly from fiber lasers, which will be significant breakthrough for mode-locked fiber lasers for extremely short pulses.

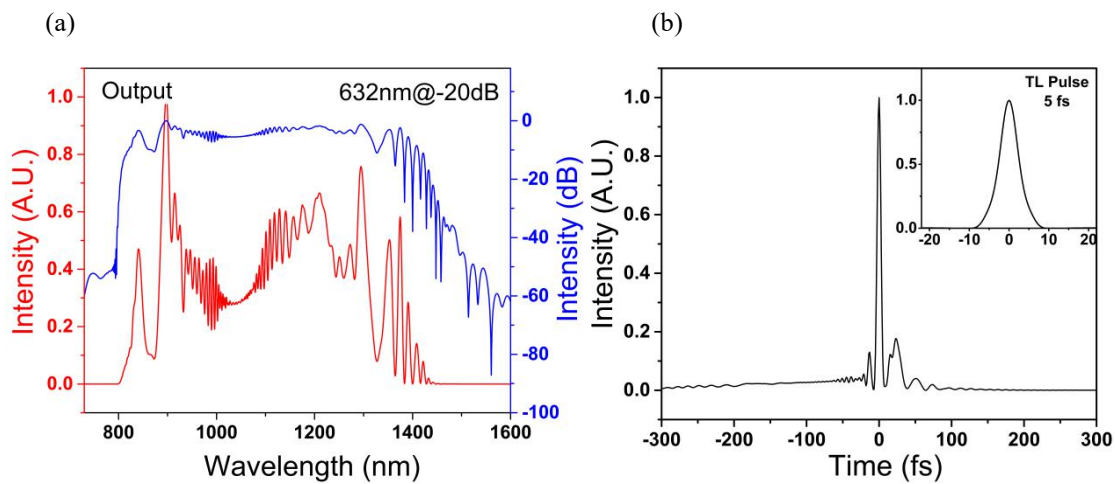

**Figure 4**. Octave spanning simulation result. (a) Output spectrum. (b) A 300 /mm grating pair dechirped pulse and its Fourier transform limited pulse.

In conclusion, we demonstrate ultra-broadband, few cycle pulses directly from fiber lasers. The extreme pulse evolution inside the cavity can be stabilized by the self-similar evolution and the perfect saturable absorber of the Mamyshev oscillator. To our best knowledge, this is the broadest spectrum and shortest pulse (17 fs) directly generate from a mode-lock fiber laser.

## Methods

**Propagation model.**

To confirm the nature of mode-locking mechanism, numerical simulations for pulse propagation in the laser cavity described by cubic complex Ginzburg-Landau equation as:

$$\frac{\partial U}{\partial t} = -i\frac{\beta_2}{2}\frac{\partial^2 U}{\partial t^2} + \frac{\beta_3}{6}\frac{\partial^3 U}{\partial t^3} + \frac{g}{2}U + i\gamma|U|^2 U + i\gamma T_R \frac{\partial^2 U}{\partial t^2}U$$

Here, $U = U(z,t)$ is the slowly varying amplitude of the pulse envelope, z is the propagation coordinate, and t is the time delay parameter. $\beta_2$, $\beta_3$ and $\beta_4$ are the second-order (GVD), third-order dispersion, fourth-order dispersion parameters, respectively. $\gamma$ is the nonlinearity parameter given by $\gamma = n_2\omega_0/cA_{eff}$, where $n_2$ is the Kerr coefficient, $\omega_0$ the central angular frequency, $c$ the velocity of the light in vacuum, and $A_{eff}$ the effective mode area. $T_R$ is related to the slope of the Raman gain spectrum, which is assumed to vary linearity with frequency around the central frequency. The gain saturation is given by $g(E) = \frac{g_0}{1 + E/E_{Sat}}$, where $E$ is the pulse energy given by $E = \int_{-T_R/2}^{T_R/2}|A(z,t)|^2 dt$, $E_{Sat}$ is the saturation energy, $T_R$ is the cavity round trip time. All the parameters using in the simulation are shown in Table 2.

**Table.2 Parameters used in the simulation**

| Component | Length (m) | Bandwidth (nm) | Wavelength (nm) | Output Coupling | GVD (fs²/mm) | TOD (fs³/mm) | FOD (fs⁴/mm) | Loss |
|---|---|---|---|---|---|---|---|---|
| Filter1 | | 4 | 1040 | | | | | 30% |
| Coupling loss | | | | | | | | 25% |
| Passive fiber | 0.7 | | | | 22.2 | 63.8 | | |
| Gain fiber | 3.1 | 40 | 1045 | | 24.9 | 59 | | |
| Passive fiber | 0.7 | | | | 22.2 | 63.8 | | |
| PBS1 | | | | 40% | | | | |
| Filter 2 | | 4 | 1050 | | | | | 30% |
| Coupling loss | | | | | | | | 25% |
| Passive fiber | 0.7 | | | | 22.2 | 63.8 | | |
| Gain fiber | 3.1 | 40 | 1045 | | 24.9 | 59 | | |
| Passive fiber | 0.7 | | | | 22.2 | 59 | | |
| DDL | | | | | -3100 | | | |
| PCF | 0.45 | | | | 3.7 | -6.84 | 160 | |
| Output | | | | 80% | | | | |

## Acknowledgements

We acknowledge the financial support by the National Science Foundation (NSF) (ECCS 1710914) and the financial support from the program of China Scholarships Council (No. 201706170153). We also acknowledge center for photonics and photonic material of University of Bath provide the all normal disperison photonic crystal fiber for us.